\documentclass[12pt]{article}
\newcommand{\beq}{\begin{equation}\displaystyle}
\newcommand{\eeq}{\end{equation}}
\newcommand{\bit}{\begin{itemize}}
\newcommand{\eit}{\end{itemize}}
\newcommand{\ben}{\begin{enumerate}}
\newcommand{\een}{\end{enumerate}}
\newcommand{\bc}{\begin{center}}
\newcommand{\ec}{\end{center}}

\begin{document} 
\baselineskip 19pt

\bc \large \bf Pump induced Autler-Townes effect and A-T mixing in  four
level atoms \ec  \vskip 2.5cm

  \vskip 1.5cm

\noindent {\bf PACS number(s): 42.50 Hz, 42.50 Gy, 32.10. Fn}  \vskip 3.5cm

\begin{center} \bf\normalsize Dipankar Bhattacharyya
\footnote[2]{Permanent address: Santipur College, P.O.- Santipur, Nadia, W.B., INDIA},
Biswajit Ray and Pradip N. Ghosh\,\footnote[1]{Corresponding author.} \\ \,Department of
Physics, University of Calcutta \\ 92, A. P. C. Road, Calcutta --
700 009, \\ I N D I A \\ (Phone : +91 33 2350 8386, Fax : +91 33
2351 9755, \\ E--mail : png@cubmb.ernet.in) 
\end{center}

\newpage\begin{abstract}\baselineskip 22pt
It is shown by theoretical simulation that tuning of the pump power can induce
mixing and crossing of Autler-Townes(A-T) components of closely spaced transitions
in atoms. Pump radiation also leads to small shifts of the central hole of
A-T doublet. Off-resonance pumping gives an asymmetry in the A-T components and
by controlling pump frequency detuning it is also possible to mix the A-T components.
\end{abstract}

\newpage\noindent {\bf I Introduction:}

The interaction of a strong electromagnetic field with two, three and four-level
atomic systems leads to many interesting nonlinear effects like electromagnetically
induced transparency (EIT) [1,2], lasing without inversion(LWI) [3,4] and enhancement
or suppression of atomic refractive index [5,6,7].
The strong pump field leads to the dynamic(ac) Stark splitting or Autler-Townes(A-T)
splitting [8], in this case the system is said to be dressed by the strong pump field.
This energy level splitting is well resolved when the Rabi-frequency of the pump field is larger
than the atomic decay rates. Detailed studies on the splitting have been carried out by many
researchers both experimentally and theoretically [9-13].
In the weak probe limit this well known nonlinear effect has drawn
much attention in recent years after successful achievement of EIT and LWI in
various atomic vapor systems [2,10,14-15].

Interaction of atomic systems having closely spaced energy levels with
electromagnetic radiations leads to interesting effect on the line shape of the
observed spectra [16-19]. Schl$\ddot{o}$ssberg and Javan [16] first studied the interaction of a
three level system(TLS) having two closely spaced upper levels with two incident radiation
fields with two closely spaced monochromatic frequencies separated by an amount larger than
the natural line widths of the atomic resonances. Double resonance line shape arising
from the interaction of a four-level atomic system with two closely spaced
intermediate levels under the influence of a fixed pump frequency has been studied
[17]. Almost all of the theoretical and experimental work on A-T effect deal with 
the three level systems(TLS) and in a TLS 
two peaks of the A-T doublet occur at probe detunings $\delta_{1} = \frac{\delta_{2}}{2}$
${\pm} \frac{1}{2} \sqrt{{\delta_{2}}^{2} + {4}{|\chi|}^{2}}$ [10]
where $\delta_{2}$ and $\chi$ are the pump detuning and pump Rabi-frequency
respectively. If pump frequency is on-resonance ($\delta_{2}$ = 0) the two
symmetric peaks appear at ${\delta_{1}} = {\pm}{\chi}$ with identical linewidths.
For non-zero detuning, the peaks are asymmetric and one peak has
larger linewidth, while the other has smaller linewidth. This change of
linewidth occurs in such a way that the sum of the linewidths is equal to the unperturbed
linewidth of the A-T components [10].

Hyperfine transitions in alkaline atoms like Rubidium are widely used for
laser cooling [20]. Rb-D$_{2}$ hyperfine transitions of both the isotopes
involve a single lower level and three closely spaced upper levels.
Recently we reported [21] a theoretical simulation based on rate equations
of a four level atom with a common ground level and three closely spaced upper
levels interacting with a standing wave. They were solved numerically
to show the difficulty of isolating the components in Lamb dip spectroscopy.
Effect of pump Rabi frequency and detuning on A-T components of atomic system
with three closely spaced upper levels, commonly found in the hyperfine splitting
of alkali atoms, has not been studied theoretically earlier.
Here we present a perturbative analytical treatment of such a system 
with the pump frequency held near resonance 
with one of the hyperfine transitions (F$_{g}$ = 2 $\rightarrow$ F$_{e}$ = 1
for $^{85}$Rb-D$_{2}$). It leads to Autler-Townes splitting of the
two closely spaced probe transitions (F$_{g}$ = 2 $\rightarrow$ F$_{e}$ = 2 and 3).
It is found that tuning of the pump power with
on resonance pump frequency results in mixing and crossing of
the A-T components in $^{85}Rb$. At higher pump power the mixing of the
A-T components leads to a significant outward shift in the holes created by
the overlap of the A-T doublets. Effect of off-resonance pumping on the
line shape of the A-T components is also discussed. 

\noindent {\bf II Theory:} 

The four-level system (Fig.\,1) with energy levels E$_{1}$, E$_{2}$, E$_{3}$
and E$_{4}$ interacts with two coherent monochromatic
radiation fields of frequencies ${\Omega}_{1}$ and ${\Omega}_{2}$.
The strong pump frequency ${\Omega}_{1}$ is nearly equal to
transition frequency ${{\omega}_{1}}
= {\frac{(E_{2} - E_{1})}{\hbar}}$,
and the field ${{\Omega}_{2}}$ probes the
transitions 1 ${\rightarrow}$ 3 and 1 ${\rightarrow}$
4 with the transition frequencies ${{\omega}_{2}} =
{\frac{(E_{3} - E_{1})}{\hbar}}$ and 
${{\omega}_{3}} = {\frac{(E_{4} - E_{1})}{\hbar}}$
respectively.

The Hamiltonian of the system can be written as
\begin{eqnarray}
                  H = {H}_{0} + {H}_{I}
\end{eqnarray}
where ${{H}_{0}}$ is the unperturbed atomic Hamiltonian with the
eigenvalues E$_{1}$, E$_{2}$, E$_{3}$ and E$_{4}$. 

${{H}_{I}}$ is the interacting Hamiltonian of the four-level system and
can be written as
\begin{eqnarray}
{H}_{I} = -2{\varepsilon}_{1}{\mu} \cos{{\Omega}_{1}}{t} - 2{\varepsilon}_{2}{\mu} \cos{{\Omega}_{2}}{t}
\end{eqnarray}
where the ${\varepsilon}_{1}$ and ${\varepsilon}_{2}$ are the electric-field
amplitudes of the pump and probe radiations respectively and $\mu$ is the transition
dipole moment. In a semiclassical treatment
for the atom-field interaction we define the Rabi-frequencies as
${{\chi}_{1}}$  = ${\frac{{2}{\mu}_{12}{\varepsilon}_{1}}{\hbar}}$,  
${{\chi}_{2}}$  = ${\frac{{2}{\mu}_{13}{\varepsilon}_{2}}{\hbar}}$ 
and ${{\chi}_{3}}$  = ${\frac{{2}{\mu}_{14}{\varepsilon}_{2}}{\hbar}}$. 

The time evolution of the density matrix is given by [22,23],
\begin{eqnarray}
{\it{i} \hbar \frac{{\partial}{\rho}}{{\partial}{t}}} = [H,{\rho}] + (relaxation terms) 
\end{eqnarray}

The detunings are defined as
$\Delta{\omega}_{1}$ = (${{\Omega}_{1}}$ - ${{\omega}_{1}}$),
${\Delta}{{\omega}_{2}}$ = (${{\Omega}_{2}}$ - ${{\omega}_{2}}$) and 
${\Delta}{{\omega}_{3}}$ = (${{\Omega}_{2}}$  -  ${{\omega}_{3}}$)
and the energy level differences are
${\Delta}_{1}$ = ${\frac{({E}_{3}  - {E}_{2})}{\hbar}}$, 
${\Delta}_{2}$ = ${\frac{({E}_{4}  - {E}_{2})}{\hbar}}$ 
and ${\Delta}_{3}$ = ${\frac{({E}_{4}  - {E}_{3})}{\hbar}}$.
We introduce ${T}_{1}$, ${T}_{2}$ as the longitudinal and
transverse relaxation constants. 
Under the effect of the applied field a macroscopic polarization
will be developed in the system, which can be written as 
\begin{eqnarray}
P = N[{\mu_{21}} {\rho_{12}} \exp(\it{i}{\Omega}_{1}t) +
{\mu_{31}} {\rho_{13}} \exp(\it{i}{\Omega}_{2}t) +
{\mu_{41}} {\rho_{14}} \exp(\it{i}{\Omega}_{2}t) + c.c ].  
\end{eqnarray}

where N is the number of atoms per unit volume.
This allows us to introduce the real and imaginary parts of the pump and probe polarizations
associated with the three allowed transitions, given by
${{P}_{pr}}$ + ${\it{i}}{{P}_{pi}}$ = ${N}{{\mu}_{21}}{{\rho}_{12}}$, 
${{P}_{sr1}}$ + ${\it{i}}{{P}_{si1}}$ = ${N}{{\mu}_{31}}{{\rho}_{13}}$ 
and ${{P}_{sr2}}$ + ${\it{i}}{{P}_{si2}}$ = ${N}{{\mu}_{41}}{{\rho}_{14}}$. 
The population differences are
${\Delta}{N_{1}}$ = ${N}({{\rho}_{11}} - {{\rho}_{22}})$, 
${\Delta}{N_{2}}$ = ${N}({{\rho}_{11}} - {{\rho}_{33}})$, 
and ${\Delta}{N_{3}}$ = ${N}({{\rho}_{11}} - {{\rho}_{44}})$. 
They relax to ${{\Delta}{N_{1}}}^{o}$, ${{\Delta}{N_{2}}}^{o}$ and
${{\Delta}{N_{3}}}^{o}$ with relaxation time T$_{1}$.
The transitions 2 ${\rightleftharpoons}$ 3, 2 ${\rightleftharpoons}$ 4
and 3 ${\rightleftharpoons}$ 4 are not induced by any radiation
but ${{\rho}_{23}}$, ${{\rho}_{24}}$ and
${{\rho}_{34}}$ are not equal to zero. This will involve the non-linear terms given by
${{P}_{nr}}$ + ${\it{i}}{{P}_{ni}}$ = ${N}{{\rho}_{23}}{{\mu}_{31}} $, 
${{P}_{mr}}$ + ${\it{i}}{{P}_{mi}}$ = ${N}{{\mu}_{41}}{{\mu}_{12}}{{\rho}_{24}}$ 
and ${{P}_{qr}}$ + ${\it{i}}{{P}_{qi}}$ = ${N}{{\mu}_{41}}{{\mu}_{13}}{{\rho}_{34}}$.
The physical significance of the terms ${{P}_{nr}}$, ${{P}_{ni}}$, ${{P}_{mr}}$,
${{P}_{mi}}$, ${{P}_{qr}}$ and ${{P}_{qi}}$ was described earlier[17-19].
Fifteen steady-state optical Bloch
equations can be written in terms of polarizations, population differences and
the non-linear parameters.

\noindent{\bf III Line Shape of the Probe transitions:}

The analytical expression of the total probe polarization, 
is the sum of ${P_{si1}}$ and ${P_{si2}}$.
The analytical closed forms of the ${P_{si1}}$ and ${P_{si2}}$ are very complicated,
and to get a simplified form we consider the relaxation
parameters ${T}_{1}$ = ${T}_{2}$ = ${T}$ and probe Rabi-frequencies
${\chi_{2}}$ = ${\chi_{3}}$ = ${\chi_{p}}$.
Since pump electric field amplitude is much greater than the probe electric
field amplitude; i.e., ${\varepsilon}_{1}$ $>>$ ${\varepsilon}_{2}$
we get ${\chi_{1}}$ $>>$ ${\chi_{p}}$.

The probe polarization in zeroth order is

\begin{eqnarray}
P_{probe}  = - \Bigg[ \frac{C_{1}}{ 1 +
{\Delta\omega_{2}}^{2}{T^{2}} + \frac{{\chi_{1}}^{2}T^{2}}{4} \left( 1
- \frac{(2 {\Delta{\omega}_{2}}
- {\Delta{\omega}_{1}})^{2}{T}} {\frac{1}{T} + ({\Delta{\omega}_{1}}-
{\Delta{\omega}_{2}})^{2}{T} + \frac{T{\chi_{1}}^{2}}{4}} \right)} + \\ 
\frac{C_{2}}{1 +
{\Delta{\omega}_{3}}^{2}T^{2} + \frac{{\chi_{1}}^{2}T^{2}}{4}
\left( 1 - \frac{(2 {\Delta{\omega}_{3}}
- {\Delta{\omega}_{1}})^{2}{T}} {\frac{1}{T} + ({\Delta{\omega}_{1}}-
{\Delta{\omega}_{3}})^{2}{T} + \frac{T{\chi_{1}}^{2}}{4}} \right)} \Bigg] \nonumber
\end{eqnarray}

The parameters ${C}_{1}$ and ${C}_{2}$ are proportional to 
${{\Delta}{N_{2}}^{0}} {{\mu}_{13}}$ and ${{\Delta}{N_{3}}^{0}} {{\mu}_{14}}$ respectively.
The first and second terms of eq.(5) correspond
to the Lorentzian line shapes of 1 $\rightarrow$ 3 and 1 $\rightarrow$ 4
transitions respectively.

When pump frequency is on-resonance with the transition 1 $\rightarrow$ 2
having transition frequency ${\omega}_{1}$, the corresponding detuning is equal
to zero(i.e. ${\Delta{\omega_{1}}}$ = 0). In that condition the first term of the
eq.(5) becomes

\begin{eqnarray}
P_{probe}(\Delta{\omega}_{2})  = -\Bigg[ \frac{C_{1}}{1 + 
{\Delta{\omega}_{2}}^{2}{{T}^{2}} + \frac{{\chi_{1}}^{2}{T}^{2}}{4}\left( 1
- \frac{{4}{\Delta{\omega}_{2}}^{2}{T}}{\frac{1}{T} + 
{\Delta{\omega}_{2}}^{2}{T} + \frac{T}{4}{{\chi_{1}}}^{2}}\right)}\Bigg]  
\end{eqnarray}

Eq.(6) gives a Lorentzian line shape centered at ${\Delta{\omega}_{2}}$ = 0,
when we increase the power of pump-beam
it is split into two symmetrical components which are commonly known as A-T
doublet [8]. Similar result will be obtained from the second term of
eq.(5). The separation of the A-T components increases with increase in the value
of {$\chi_{1}$$T$}. They are not observed at lower value of {${\chi_{1}}T$}.

To see the effect of probe power on absorptive line we also present here
the expression of ${P_{probe}}$ after retaining up to second order
term of the probe Rabi-frequencies $\chi_{p}$

\begin{eqnarray}
P_{probe}  = - \left[ \frac{C_{1}}{ a_{1} + b_{1} - c_{1} - d_{1} - e_{1}}
+ \frac{C_{2}}{ a_{2} + b_{2} - c_{2} - d_{2} - e_{2}} \right] 
\end{eqnarray}

Where the above parameters $a_{1}$, $b_{1}$, $c_{1}$, $d_{1}$ and $e_{1}$ are
defined as
\begin{eqnarray*}
a_{1} = {1 + T^{2}
\left[{\chi_{p}}^{2} + \frac{1}{4}({\chi_{1}}^{2} + {\chi_{p}}^{2}) +
{\Delta{\omega}_{2}}^{2} \right]}
\end{eqnarray*}

\begin{eqnarray*}
b_{1} = \frac{({2}\Delta{{\omega}_{2}} - {\Delta{\omega}_{3}}
)({2}{\Delta{\omega}_{2}} - {\Delta{\omega}_{1}})
{{\chi_{1}}^{2}{\chi_{p}}^{2}}{T^{4}}}{8[\frac{1}{T} + (\Delta{{\omega}_{1}} -
{\Delta{\omega}_{2}})^{2}T
+ \frac{T}{4}({\chi_{1}}^{2} + {\chi_{p}}^{2})]
[\frac{1}{T} + {(\Delta{{\omega}_{2}} -
{\Delta{\omega}_{3}})}^{2}T +
\frac{T}{2}{\chi_{p}}^{2}]}
\end{eqnarray*}

\begin{eqnarray*}
c_{1} = \frac{(2\Delta{\omega}_{2}
- \Delta{\omega}_{1})^{2}{\chi_{1}}^{2}T^{3}}
{4[\frac{1}{T}
+ (\Delta{\omega}_{1} - \Delta{\omega}_{2})^{2}T
+ \frac{T}{4}({\chi_{1}}^{2} + {\chi_{p}}^{2})]} 
\end{eqnarray*}

\begin{eqnarray*}
d_{1} = \frac{(2\Delta{\omega}_{2}
- \Delta{\omega}_{3})^{2}{\chi_{p}}^{2}T^{3}}{4[\frac{1}{T} +
(\Delta{\omega}_{2} - \Delta{\omega}_{3})^{2}T
+ \frac{T}{2}{\chi_{p}}^{2}]}
\end{eqnarray*}

\begin{eqnarray*}
\hspace{-2.0cm}{
e_{1} =  \frac{{\chi_{1}}^{2}{\chi_{p}}^{2}T^{4}
 \left[ 3 + \frac{(2\Delta{\omega_{2}} - \Delta{\omega_{1}})
(2\Delta{\omega_{1}}
- \Delta{\omega_{2}})T}{\frac{1}{T} + (\Delta{\omega_{1}} - \Delta{\omega_{2}}
)^{2}T + \frac{T}{4}({\chi_{1}}^{2}
+ {\chi_{p}}^{2})}\right]^{2}}{16\left[ 1 + T^{2}({\chi_{1}}^{2} + \frac{{\chi_{p}}^{2}}{2} + {\Delta{\omega}_{1}}^{2}) 
- \frac{T^{2}}{4}
\left( \frac{{\chi_{p}}^{2}(2\Delta{\omega_{1}} 
 - \Delta{\omega}_{3})^{2}}{\frac{1}{T} +
(\Delta{\omega}_{1} - \Delta{\omega}_{2})^{2}T + \frac{T}{4}({\chi_{1}}^{2} + 
{\chi_{p}}^{2})} +
 \frac{{\chi_{p}}^{2}(2\Delta{\omega_{1}} - \Delta{\omega_{2}})^{2}}{\frac{1}
{T} +
(\Delta{\omega_{1}} -  \Delta{\omega_{2}})^{2}T + \frac{T}{4}({\chi_{1}}^{2} 
+ {\chi_{p}}^{2})}\right) \right]}}
\end{eqnarray*}

\par similarly $a_{2}$, $b_{2}$, $c_{2}$, $d_{2}$ and $e_{2}$ are

\begin{eqnarray*}
a_{2} = {1 + T^{2}
\left[{\chi_{p}}^{2} + \frac{1}{4}({\chi_{1}}^{2} + {\chi_{p}}^{2}) +
{\Delta{\omega}_{3}}^{2} \right]}
\end{eqnarray*}

\begin{eqnarray*}
b_{2} = \frac{({2}\Delta{{\omega}_{3}} - {\Delta{\omega}_{1}})
({2}{\Delta{\omega}_{3}} - {\Delta{\omega}_{2}})
{{\chi_{1}}^{2}{\chi_{p}}^{2}}{T^{4}}}{8[\frac{1}{T} + (\Delta{{\omega}_{1}} -
{\Delta{\omega}_{3}})^{2}T
+ \frac{T}{4}({\chi_{1}}^{2} + {\chi_{p}}^{2})]
[\frac{1}{T} + {(\Delta{{\omega}_{2}} -
{\Delta{\omega}_{3}})}^{2}T +
\frac{T}{2}{\chi_{p}}^{2}]}
\end{eqnarray*}

\begin{eqnarray*}
c_{2} = \frac{(2\Delta{\omega}_{3}
- \Delta{\omega}_{1})^{2}{\chi_{1}}^{2}T^{3}}
{4[\frac{1}{T}
+ (\Delta{\omega}_{1} - \Delta{\omega}_{3})^{2}T
+ \frac{T}{4}({\chi_{1}}^{2} + {\chi_{p}}^{2})]} 
\end{eqnarray*}

\begin{eqnarray*}
d_{2} = \frac{(2\Delta{\omega}_{3}
- \Delta{\omega}_{2})^{2}{\chi_{p}}^{2}T^{3}}{4[\frac{1}{T} +
(\Delta{\omega}_{2} - \Delta{\omega}_{3})^{2}T
+ \frac{T}{2}{\chi_{p}}^{2}]}
\end{eqnarray*}

\begin{eqnarray*}
\hspace{-2.0cm}{
e_{2} =  \frac{{\chi_{1}}^{2}{\chi_{p}}^{2}T^{4}
 \left[ 3 + \frac{(2\Delta{\omega_{1}} - \Delta{\omega_{3}})
(2\Delta{\omega_{3}}
- \Delta{\omega_{1}})T}{\frac{1}{T} + (\Delta{\omega_{1}} - \Delta{\omega_{2}}
)^{2}T + \frac{T}{4}({\chi_{1}}^{2}
+ {\chi_{p}}^{2})}\right]^{2}}{16\left[ 1 + T^{2}({\chi_{1}}^{2} + \frac{{\chi_{p}}^{2}}{2} + {\Delta{\omega}_{1}}^{2}) 
- \frac{T^{2}}{4}
\left( \frac{{\chi_{p}}^{2}(2\Delta{\omega_{1}} 
 - \Delta{\omega}_{3})^{2}}{\frac{1}{T} +
(\Delta{\omega}_{1} - \Delta{\omega}_{2})^{2}T + \frac{T}{4}({\chi_{1}}^{2} + 
{\chi_{p}}^{2})} +
 \frac{{\chi_{p}}^{2}(2\Delta{\omega_{1}} - \Delta{\omega_{2}})^{2}}{\frac{1}
{T} +
(\Delta{\omega_{1}} -  \Delta{\omega_{2}})^{2}T + \frac{T}{4}({\chi_{1}}^{2} 
+ {\chi_{p}}^{2})}\right) \right]}}
\end{eqnarray*}

\newpage \noindent{\bf IV Computation of line shape in Rubidium atom:}

For the purpose of numerical calculations we have chosen a four-level system that corresponding to
$^{85}$Rb$-$$D_{2}$ hyperfine transitions (Fig.\,1).
The lower energy level 1 corresponds to the $F_{g}$ $=$ $2$
hyperfine level in $^{5}$${S}_{\frac{1}{2}}$ ground state of $^{85}$${Rb}$$-$$D_{2}$,
while the upper levels 2, 3 and 4 are the $F_{e}$ = 1, 2, 3 components of
the $^{5}$${P}_{\frac{3}{2}}$ excited state.
The energy level differences in the upper states of $^{85}$Rb are ${{\Delta}_{1}}$ = 0.029$GHz$,
${{\Delta}_{2}}$ = 0.092$GHz$ and ${{\Delta}_{3}}$ = 0.063$GHz$ respectively.
The value of the pump frequency used in this work is
384615.38 $GHz$(${\lambda \approx 780.0}nm$).
The values of pump Rabi-frequencies used for computation are very close to
the experimental situation
of $Rb$ vapor system [11,24]. We choose the value of relaxation constants
$\frac{1}{T}$ = ${\Gamma}$ = 0.01$GHz$ [11].

     Fig.\,2A shows the probe polarization vs.\ detuning $\left( \Omega_{2} -
 \frac{(\omega_{2} + \omega_{3})}{2} \right)$ curve, for two different 
pump powers when the pump frequency is held on-resonance
with the transition 1 $\rightarrow$ 2.
The pump field Rabi-frequencies used in Fig.\,2A
are 0.01 and 0.04 $GHz$.
At low pump Rabi-frequency two absorption lines appear corresponding to
1 $\rightarrow$ 3 and 1 $\rightarrow$ 4 transitions (Fig.\,2A(a))
when the probe is tuned across the lines.
As we increase the pump Rabi-frequency keeping
${\Gamma}$ constant the two absorption lines become power broadened
and each line is split into two A-T components (Fig.\,2A(b)) denoted as
(A$_{1}$, A$_{2}$) and (B$_{1}$, B$_{2}$). 

It is observed that at a pump Rabi-frequency of 0.06$GHz$ the nearby split
components of two A-T doublets corresponding to the transitions 1 $\rightarrow$ 3 
and 1 $\rightarrow$ 4 are mixed with each other (A$_{2}$ and B$_{1}$ of Fig.\,2B).
This mixing occurs at a frequency 
$\frac{({\omega_{2}} + {\omega_{3}})}{2}$ and forms a single stronger absorption line. 
Thus one can obtain three A-T components instead of four (Fig.\,2B(a)).
In order to find the effect of the finite probe power the probe
polarization(eq.7) is plotted in Fig.\,2B(b),
the values of $\chi_{1}$ and $\Gamma$ remain unchanged; value of
$\chi_{p}$ is 0.006$GHz$. The effect of the finite probe power
is revealed as a small change of line shape.

In Fig.\,2C we see the effect of further higher pump powers on the absorption line.
At pump Rabi-frequency 0.1$GHz$ the central line of Fig.\,2B is
again split and they cross the zero detuning line (Fig.\,2C(a)).
With further increase in pump power the separation between two A-T components
increases but they never mix again. As a consequence positions of the
holes produced by the A-T splitting also shifts with pump power,
this is observed after they cross the zero probe detuning.

In Fig.\,3 we present two spectra (A,B) where we show the effect of
off-resonance pumping on the probe absorption. In both the spectra (A,B) dotted curve(a)
shows the probe absorption when $\Delta{\omega_{1}}$ = 0 and $\chi_{1}$ = 0.03$GHz$
and it leads to the two A-T doublets corresponding to two allowed transitions.
The solid curve(b) of Fig.\,3A plotted with $\chi_{1}$ = 0.03 and
$\Delta{\omega_{1}}$ = -0.012 $GHz$ gives four A-T peaks with asymmetric linewidths
because of the non-zero detuning of the pump field. A-T components A$_{1}$ and
B$_{1}$ are weaker and have lower linewidth than A$_{2}$ and B$_{2}$. 
We can control the pump detuning in such a way that at a value of -0.026$GHz$
the A-T component(B$_{1}$) of 1 $\rightarrow$ 4 transition nearly mixes with
the A-T component(A$_{2}$) of the 1 $\rightarrow$ 3 transition (Fig.\,3B) solid
curve(b). So by controlling the pump detuning (i.e. coherent control) we are able
to mix the nearby A-T component keeping pump power fixed.
In both the cases mentioned above the sum of the linewidths
of two A-T components is equal to the unperturbed linewidth [10] for each transition.
                      
\noindent {\bf V Conclusions:}

The power-broadened A-T doublets in a four-level system interacting
with a near resonant strong pump and a weak probe field have been investigated
by analytical solution of the optical Bloch equations. 
Numerical computation of the line shape of $^{85}$Rb-D$_{2}$ transitions is
reported. In a four level system two A-T doublets are generally expected
in the presence of a pump.
Our calculation of the A-T splitting in a four-level system with three closely
spaced upper levels reveals that the pump power can be tuned so that the
nearby A-T components of two nearby transitions mix with each other.
Such mixing of the A-T components may also be achived by varying detuning of the pump frequency.
Further increase of pump power will cause the mixed components to split again
and they cross to the other sides of the mean frequency resulting in four components.
It is noteworthy that the positions of the holes of the A-T doublets shifts
with increase of pump Rabi-frequency. 
This results from increased A-T splitting.
So one can manipulate the atomic response through pump field
either by pump power or by pump detuning.
We also performed the computation with another $^{85}$Rb-D$_{2}$ transitions
(i.e. $F_{g}$ = 3 $\rightarrow$ $F_{e}$ = 2, 3, 4) where the upper state hyperfine
energy level spacings are higher than there of the $^{85}$Rb-D$_{2}$ transitions
corresponding to $F_{g}$ = 2 $\rightarrow$ $F_{e}$ = 1, 2, 3. 
In this case much higher pump power is required for the level mixing.

\noindent{\bf Acknowledgement:}
Authors thank the Department of Atomic Energy Government of India for 
award of a research grant.

\noindent {\bf References:}
\ben
\setlength{\itemsep}{0ex plus0.2ex}
\setlength{\parsep}{0.5ex plus0.2ex minus0.1ex}
\baselineskip 24pt
\item J. E. Field, K. H. Hahn and S. E. Harris 1991 Phys. Rev. Lett. {\bf 67} 3062.
\item J. Boller, A. Imamoglu and S. E. Harris 1991 Phys. Rev. Lett. {\bf 66} 2593.
\item S. E. Harris, 1989 Phys. Rev. Lett. {\bf 62} 1033.
\item M. O. Scully 1992 Phys. Rep. {\bf 219} 191.
\item M. O. Scully 1991 Phys. Rev. Lett. {\bf 67} 1855.
\item M. O. Scully and M. Fleischhauer 1992 Phys. Rev. Lett. {\bf 69} 1360.
\item S. E. Harris 1994 Opt. Lett. {\bf 19} 2018.
\item S. H. Autler and C. H. Townes 1955 Phys. Rev. {\bf 100} 703.
\item Y. Zhu 1993 Phys. Rev. A. {\bf 47} 495.
\item G. Vemuri, G. S. Agarwal and B. D. Nageswara Rao 1996 Phys. Rev. A. {\bf 53} 2842.
\item Y. Zhu and T. N. Wasserluf 1996 Phys. Rev. A. {\bf 54} 3653.
\item C. Wei, N. B. Manson and John P. D. Martin 1995 Phys. Rev. A. {\bf 51} 1438.
\item D. Mc Gloin 2003 J. Phys. B: At. Mol. Opt. Phys. {\bf 36} 2861.
\item A. S. Zibrov, M. D. Lukin, D. E. Nikonov, L. Hollberg, M. O. Scully,
      V. L. Velichansky and H. G. Robinson 1995 Phys. Rev. Lett. {\bf 75} 1499.
\item G. G. Padmabandu, G. R. Welch, I. N. Shubin, E. S. Fry, D. E. Nikonov,
      M. D. Lukin and M. O. Scually 1996 Phys. Rev. Lett. {\bf 76} 2053.
\item H. R. Schl$\ddot{o}$sberg and A. Javan 1966 Phys. Rev. {\bf 150} 267.
\item S. Mandal and P. N. Ghosh 1993 Phys. Rev. A. {\bf 47} 4934.
\item S. Mandal and P. N. Ghosh 1992 Phys. Rev. A. {\bf 45} 4990.
\item P. N. Ghosh and S. Mandal 1989 Chem. Phys. Lett. {\bf 64} 279.
\item C. E. Wieman, G. Flowers and D. Gilbert 1995 Am. J. Physics. {\bf 63} 317.
\item D. Bhattacharyya, B. K. Dutta, B. Ray and P. N. Ghosh 2004 Chem. Phys. Lett. {\bf 389} 113.
\item S. Stenholm, "Foundations of Laser Spectroscopy" (John Willey, New York) 1983.
\item "Quantum Optics" by M. O. Scully and M. S. Zubairy, (Cambridge University Press London 1997).
\item M. Yan, E. G. Rickey and Y. Zhu 2001 Opt. Lett. {\bf 26} 548.
\een

\newpage
\thispagestyle{empty}
\noindent{\bf $\:\,$Figure Captions}
\newcounter{fig}
\setlength{\labelwidth}{2.cm} 
\setlength{\leftmargin}{1.37cm}
\setlength{\labelsep}{0.25cm} 
\setlength{\rightmargin}{0.1mm}
\setlength{\parsep}{0.5ex plus0.2ex minus0.1ex}
\setlength{\itemsep}{0ex plus0.2ex} \baselineskip 21pt

\noindent {\bf{Figure  1:}}
      Energy level diagram of a four-level atom interacting with two radiation
      fields. Inset: the Energy-level of $^{85}{Rb} - {D}_{2}$ transitions.

\noindent {\bf{Figure  2A:}}
      Probe polarization P$_{probe}$ vs.\
      $\left( \Omega_{2} - \frac{(\omega_{2} + \omega_{3})}{2} \right)$ curve
      for (a) ${\chi_{1}}$ = 0.01 and (b) ${\chi_{1}}$ = 0.04 $GHz$. 
      The pump field is on-resonance with 1 $\rightarrow$ 2 level.      
      The parameters used in calculations are $\Gamma$ = 0.01 $GHz$ and
      ${\omega_{1}}$ = 384615.38 $GHz$
                
\noindent {\bf{Figure  2B:}}
      Probe polarization P$_{probe}$ vs.\
      $\left( \Omega_{2} - \frac{(\omega_{2} + \omega_{3})}{2} \right)$ curve.
      (a) ${\chi_{1}}$ = 0.06 $GHz$. The curve(b) based on eq.(7) uses
      ${\chi_{1}}$ = 0.06 and ${\chi_{p}}$ = 0.006 $GHz$. Other parameters 
      remain the same as in Fig.\,2A. 
 
\noindent {\bf{Figure  2C:}}
      Probe polarization P$_{probe}$ vs.\
      $\left( \Omega_{2} - \frac{(\omega_{2} + \omega_{3})}{2} \right)$ curve 
      for (a) ${\chi_{1}}$ = 0.1 and (b) ${\chi_{1}}$ = 0.12 $GHz$. 
      Other parameters remain the same as in Fig.\,2A.

\noindent{\bf{Figure  3:}}
      Probe polarization P$_{probe}$ vs.\
      $\left( \Omega_{2} - \frac{(\omega_{2} + \omega_{3})}{2} \right)$ curve 
      for figure (A) and (B)  the dotted curve (a) ${\chi_{1}}$ = 0.03 $GHz$
      and $\Delta{\omega_{1}}$ = 0. In figure (A) for (b) ${\chi_{1}}$ = 0.03 and
      $\Delta{\omega_{1}}$ = - 0.012 $GHz$. In figure (B) for (b) ${\chi_{1}}$ = 0.03 and
      $\Delta{\omega_{1}}$ = - 0.026 $GHz$. Other parameters remain the same as in
      Fig.\,2A.
             
\end{document}